# Vanishing Altermagnetism and Emergent Ferromagnetism at the Two-Dimensional Limit of CrSb

Yunlong Liu[1†], Zixuan Wu[2†], Qinxi Liu[3,4†], Yupeng Zhi[3,4], Mingqiang Ren[5], Xue Jiang[3,4*], Canli Song[5], Ding Zhang[5], Xucun Ma[5], Huimin Zhang[1,2*], Jijun Zhao[3,4]

[1]Key Laboratory of Materials Modification by Laser, Ion and Electron Beams (Dalian University of Technology), Ministry of Education, Dalian 116024, China

[2]DUT-BSU Joint Institute, Dalian University of Technology, Dalian 116024, China

[3]Guangdong Basic Research Center of Excellence for Structure and Fundamental Interactions of Matter, Guangdong Provincial Key Laboratory of Quantum Engineering and Quantum Materials, School of Physics, South China Normal University, Guangzhou 510006, China

[4]School of Physics, South China Normal University, Guangzhou 510631, China

[5]State Key Laboratory of Low-Dimensional Quantum Physics, Department of Physics, Tsinghua University, Beijing 100084, China

[†]These authors contributed equally: Yunlong Liu, Zixuan Wu, Qinxi Liu

*****Correspondence to: huiminzhang@dlut.edu.cn, jiangx@scnu.edu.cn

Lead Contact's email address: huiminzhang@dlut.edu.cn

## Summary

Altermagnets offer a promising route to spintronic devices by combining spin-split electronic states with zero net magnetization, but how reduced dimensionality reshapes their magnetic order remains poorly understood. Here, we grow bilayer-thickness CrSb (0001) films directly on $SrTiO_3$(001) substrates by molecular beam epitaxy and uncover an unexpected thickness-dependent magnetic transition. The one-bilayer films exhibit room-temperature ferromagnetism with strong perpendicular magnetic anisotropy, whereas ferromagnetism vanishes in two-bilayer films, contrasting sharply with the altermagnetic ground state of bulk CrSb. Density functional theory calculations reveal that the exposed Cr surface induces orbital rearrangement that strengthens intralayer ferromagnetic exchange while suppressing competing interlayer antiferromagnetic coupling, driving the dimensional crossover from altermagnetism to ferromagnetism. These findings establish dimensionality and surface termination as powerful controls of magnetic order and provide a platform for engineering ultrathin magnetic states for spintronic applications.

**Keywords**: altermagnet; two-dimensional limit; room-temperature ferromagnetism; CrSb films; molecular beam epitaxy; density functional theory

## INTRODUCTION

Altermagnets, a newly identified magnetic ground state, are characterized by the coexistence of compensated antiferromagnetic ordering and spin-split electronic bands in reciprocal space[1,2]. Their exceptional advantages originate from a unique combination of tunable spin-splitting enabling substantial spin-polarized currents, high frequency spin dynamics, immunity to demagnetization effects, and compatibility with low-power manipulation, addressing key limitations of traditional ferromagnets and antiferromagnets for next-generation electronics[1–4]. Hallmarks of altermagnetism have been experimentally validated, including the lifting of spin degeneracy along low-symmetry trajectories within the Brillouin zone, as directly detected by angle-resolved photoemission spectroscopy (ARPES)[5,6]. Transport measurements further revealed emergent phenomena including the spontaneous anomalous Hall effect[7], Nernst effects[8,9], and non-relativistic charge-spin interconversion[4,10,11], all of which are rooted in the unique altermagnetic ordering.

Extensive theoretical and experimental efforts have been dedicated to exploring unconventional altermagnetism in emerging material systems, encompassing $\alpha$-MnTe[5,12], CrSb[6,13], $MnTe_2$[14–16], and $RuO_2$[17–19]. Altermagnetism has been revealed in layered van der Waals materials such as $CoNb_4Se_8$[20], $KV_2Se_2O$[21] and $Rb_{1-\delta}V_2Te_2O$[22], which facilitates heterostructure engineering. Recent theoretical work further suggests that the altermagnetic proximity effect can introduce spin splitting in neighboring layers[23]. Among these altermagnet materials, CrSb stands out owing to its giant spin splitting of 0.6-1.0 eV with bulk *g*-wave symmetry[1,6,24,25] and an exceptionally high Néel temperature ($T_N$ = 710 K)[26]. Prior work demonstrated the reorientation of altermagnetic order from a collinear Néel vector to a canted configuration via lattice engineering in ∼30 nm CrSb films, underscoring the critical role of spin-lattice coupling[27]. Recent *in vacuo* ARPES measurements reveals a bulk-type, three-dimensional momentum-dependent band splitting of up to 0.7 eV with *g*-wave symmetry, persisting down to 10 nm thickness[28]. Spin-resolved ARPES on 10 nm CrSb films further revealed a transition to a spin-degenerate ground state throughout the entire Brillouin zone, quantitatively demonstrating strain-controlled tuning of altermagnetic spin splitting[29]. A recent study further reported a thickness-driven magnetic transition from a ferrimagnetic state in single-unit-cell films to an altermagnetic state above the critical threshold of 7/4 unit cells in CrSb films[30].

Despite these advances, direct epitaxial growth of uniform, atomically flat CrSb films on semiconducting substrates remains unrealized, and macroscopic detection of intrinsic magnetism at the two-dimensional limit still remains elusive. Investigating the intrinsic magnetic properties of 2D CrSb films poses considerable challenges, primarily owing to the extreme difficulty in synthesizing large-area, high-quality films with uniform atomic thickness and pristine interfaces. Precise thickness modulation of CrSb thin films has long been hindered by its intrinsic three-dimensional crystal structure, which is characterized by strong interlayer coupling and abundant dangling bonds across atomic layers. Here, we exploit the strong interfacial coupling between CrSb(0001) and $SrTiO_3$(STO)(001) substrates, which stabilizes the epitaxial lattice structure and facilitates interfacial charge transfer. We achieve the epitaxial growth of bilayer CrSb films on STO(001) via molecular beam epitaxy (MBE). By combining in-situ low-temperature scanning tunneling microscopy/spectroscopy (STM/S), ex-situ macroscopic magnetic measurements, and density-functional theory (DFT) calculations, we unambiguously demonstrate the emergence of robust ferromagnetism in ultrathin CrSb/STO films.

## RESULTS

### Epitaxial growth of CrSb films on $SrTiO_3$ substrate and surface morphology

CrSb crystallizes in the hexagonal NiAs-type structure (space group $P6_3/mmc$) with lattice constants $a$ = 4.02 Å and $c$ = 5.89 Å. One unit-cell CrSb contain four alternative Sb or Cr atom layers in $c$ direction, and the electron spins of Cr atoms are ferromagnetically aligned perpendicular to the basal plane and antiferromagnetically ordered along the $c$ axis[1,6,13] (Figure S1). The opposing spin sublattices ($Cr_A$ and $Cr_B$) are not related by simple space inversion or translation, but rather by six-fold rotation or mirror symmetry operations, resulting in alternating spin polarizations. We epitaxially grew uniform CrSb films on STO(001) substrates by MBE, with experimental details provided in the Methods section. The surface morphology of a representative CrSb film is shown in Figure 1A. A zoom-in STM image (Figure 1B) reveals the characteristic morphological features of the CrSb film. The step height of the thickness unit is approximately 0.98 nm (Figure 1C), corresponding to eight Cr/Sb atomic layers along the $c$ axis, namely a bilayer structure. Hereafter, we define this fundamental epitaxial thickness unit as one bilayer (BL).

As the coverage increases from 0.83 BL (Figure 1A) to 1.0 BL (Figure 1D), the film maintains a layer-by-layer growth mode with a constant thickness of 1 BL and progressively increasing surface coverage. Beyond 1 BL, however, the film transitions to an island-like growth mode (Figure S2) and tends to develop multiple crystalline phases (Figure S3). These growth characteristics indicate a strong interfacial interaction between the 1 BL CrSb film and the STO substrate. Notably, the thickness evolution of CrSb/STO films contrasts with that of CrSb/$Bi_2Te_3$[30], which exhibits a fractional unit-cell thickness. This deviation likely arises from the distinct interfacial coupling and chemical environments that govern the growth dynamics differently at the two heterointerfaces.

Multi-domain morphology, with domain sizes on the order of tens of nanometers, is observed (Figures 1E and S4), a characteristic signature of lattice-mismatched epitaxial growth of hexagonal CrSb(0001) on tetragonal STO(001). Within each domain, well-resolved Kagome lattices are clearly visible, consisting of central hexagonal units and interconnected triangular motifs, as delineated by the dashed outline in Figure 1E. The measured Kagome lattice periodicity is 1.7 nm, corresponding to four times the in-plane lattice constant of CrSb ($a$ = 4.02 Å). Within one Kagome unit cell, pronounced high- and low-contrast features appear at the hexagonal and triangular lattice sites, respectively, yielding a height difference of approximately 0.62 Å (Figure 1F). Furthermore, the Kagome lattice structure exhibits negligible dependence on the STM bias voltage (Figure S5), indicating that it originates from an intrinsic surface Sb reconstruction, which will be discussed in detail below.

**Electronic properties of CrSb Thin Films**

To explore the electronic properties of 1BL CrSb/STO films, we performed dI/dV measurements. The CrSb/STO film exhibits a well-defined Kagome lattice, as highlighted by the overlaid dashed white Star-of-David patterns in Figure 2A. Three representative sites, sites A (hexagon center), B (triangular bridge), and C (triangle center), correspond to the reconstructed Sb, Cr, and Sb atomic sites, respectively, with site A being the farthest from the interface. Their atomic and bonding environments are illustrated in Fig. 2B. The dI/dV spectrum at site A features a prominent peak at 0.35 eV and a valence band edge near −0.7 eV (marked by red arrows in Figure 2C), with a relatively low density of states (DOS) between these two spectral features. Spectra acquired at sites B and C show overall similar line shapes but a considerably broadened peak at 0.35 eV (Figures 2D–2E). We find that the full width at half maximum (FWHM) obtained from Gaussian fits for the 1 BL CrSb film at site B (0.285 eV) is considerably broader than that at site A (0.184 eV) and site C (0.229 eV). A similar broadening is also observed for the corresponding peak in the 2 BL CrSb/STO film (Figure S6). The peak energy at site A remains nearly identical for both 1BL and 2BL CrSb films. In contrast, the corresponding peak at sites B and C shifts toward lower energy in the 2BL sample. Such thickness-dependent spectral rigid shifts suggest electron charge transfer from the CrSb film to the STO substrate. A distinct peak–dip structure emerges near the Fermi level for 1BL CrSb/STO (Figure 2F), which vanishes in 2BL CrSb/STO (Figure S7). This behavior likely originates from the complex electronic band structure inherent to the Kagome lattice[31,32].

We further performed spatially resolved dI/dV mapping on the Kagome lattice (Figure 2G with additional data are shown in Figure S8). The dI/dV maps acquired at the peak energy of ~0.4 eV above the Fermi level reveal an enhanced density of states localized at the shared triangular sites. This characteristic site-selective density of state distribution is a well-established hallmark of flat electron bands a hallmark of geometric frustration in Kagome systems[33,34]. In Kagome lattices, local uncompensated spins interact with nearly dispersionless flat bands arising from geometric frustration and destructive quantum interference, which may give rise to flat-band ferromagnetism.

**Magnetic properties of CrSb Thin Films**

The coverage of the MBE-grown CrSb film is uniform within 10% across the entire sample effective area of 6 mm × 2 mm, as confirmed by STM imaging at multiple locations (Fig. S9). This macroscopic uniformity enables reliable electrical transport and magnetization measurements on the CrSb films. For transport measurements, 1 BL CrSb films were grown on insulating STO substrates under similar growth conditions and subsequently capped with ~20 nm Si to prevent oxidation. Figure 3A presents the temperature-dependent longitudinal resistance, which decreases monotonically with decreasing temperature, indicating a typical metallic behavior with no additional features observed at low temperatures. Figure 3B displays the Hall resistance ($R_{yx}$) as a function of perpendicular magnetic field measured at various temperatures from 2 to 200 K (additional data are shown in Figure S10). Notably, a clear anomalous Hall effect (AHE) is observed, a hallmark of intrinsic ferromagnetic order in metallic systems[35]. A sign reversal of the AHE occurs between 10 and 20 K, which we attribute to interfacial effects[36–38].

For magnetic characterization, the 1 BL CrSb/STO films were capped with ~20 nm amorphous Sb (see cross-sectional TEM images in Figure S11) and measured using a magnetic property measurement system (PPMS-VSM). As shown in Figure 3C, well-defined hysteresis loops are observed in the magnetization versus magnetic field (*M–H*) curves under perpendicular magnetic fields ($H_{\perp}$). With increasing temperature from 2 to 300 K, the hysteresis loops gradually shrink, providing solid evidence for ferromagnetic ordering in the 1 BL CrSb/STO films. Notably, distinct *M–H* loops persist at 300 K, indicating that the Curie temperature exceeds room temperature. In contrast, the *M–H* loops measured under in-plane magnetic fields ($H_{//}$) exhibit negligible variation (Figure 3D), confirming that the magnetic easy axis is along the *c*-axis. Thus, perpendicular magnetic anisotropy (PMA) is unambiguously established in the 1 BL CrSb/STO films. It is worth noting that while magnetometry indicates an out-of-plane easy axis with a clear magnetic hysteresis, the AHE exhibits no such hysteresis, suggesting an out-of-plane hard axis in the transport response. This apparent discrepancy arises from the different physical quantities probed by the two techniques: magnetometry measures the volumetric magnetization of the entire sample, whereas AHE selectively probes the electrically conducting layer and is strongly affected by the interfacial effects in ultrathin films. Such decoupling between the bulk magnetic moment and the transport-derived magnetization has also been observed in other systems, including $TbMn_6Sn_6$[39,40].

Under $H_{\perp}$, the coercive field $H_c$ decreases from 397 Oe at 2 K to 215 Oe at 100 K, and further to 198 Oe at room temperature (Figure S12). The saturation magnetization $M_s$ of the bilayer CrSb/STO film is determined to be approximately $6.7 \times 10^{-4}$ emu, from which the magnetic moment per Cr atom is calculated to be 3.0 $\mu_B$ (calculations in Supplementary Note 2). These observations are reproducible across multiple 1 BL CrSb/STO samples (Figure S13). While the saturation magnetic moment exhibits slight sample-to-sample variation, a common feature of ultrathin epitaxial films due to inherent thickness and coverage fluctuations, the characteristic hysteresis loop remains fully reproducible across all measured samples. By contrast, CrSb films thicker than 1 BL show significantly suppressed magnetic hysteresis (Figure S14). The observation of anomalous Hall effect, together with the magnetization measurements, provides robust and complementary evidence for the ferromagnetic state in our CrSb/STO films.

Control experiments on uncapped 1 BL CrSb/STO films or capped STO substrate show no ferromagnetic hysteresis (Figure S14), confirming that the magnetic signal originates from the CrSb film and requires proper Sb capping to prevent oxidation. Additionally, the strict thickness dependence of the ferromagnetic signal, present only in the 1 BL film and absent in thicker films, rules out any significant contribution from $Ti^{3+}$ ions at the STO surface, as such a mechanism would not be expected to exhibit such a sharp thickness cutoff. The STO substrate merely contributes a diamagnetic background[41], which was subtracted via linear fitting procedures (Figure S15). The observed ferromagnetic signal is unambiguously attributed to the epitaxial bilayer CrSb/STO films, rather than the Sb capping layer or the STO substrate. A possible concern is that excess Cr atoms in the bilayer CrSb/STO films might contribute to the ferromagnetic signal. While Cr clusters of 2–3 nm in diameter are known to exhibit weak ferromagnetism due to structural disorder[42], our STM measurements show no evidence of such clusters. We therefore consider it highly unlikely that the observed ferromagnetic ordering originates from excess Cr atoms.

**DFT calculations on the magnetic properties of CrSb films**

We further performed DFT calculations to elucidate the structural stability and magnetic properties of 1BL CrSb/STO films. Among all the possible Cr-Sb compounds, only CrSb ($P6_3/mmc$, $a$ = 4.02 Å, $c$ = 5.89 Å) exhibits hexagonal symmetry. The thickness unit of 0.98 nm corresponds to bilayer CrSb (or eight Cr/Sb atom layers). Based on experimental observations, we construct the structural model with the in-plane crystallographic orientation relationship $[100]_{STO}//[11\text{-}20]_{CrSb}$ (Figure S16 and Supplementary Note 1). The preferential stability of the 1 BL CrSb phase arises from the strong interfacial interaction with the STO substrate (interface interaction energy: –0.11 eV/$Å^2$), which is substantially larger than that for thicker films (e.g., –0.02 eV/$Å^2$ for 2 BL). This substrate stabilization effect, combined with the positive formation energies of freestanding CrSb films, explains why the bilayer thickness is the fundamental growth unit and why integer-bilayer growth is strictly observed in our STM measurements.

To replicate the experimentally observed CrSb film surface morphology, we simulated hexagonally Sb reconstructed Cr-terminated CrSb surface (Figure 2B), which is highly plausible given the Sb-rich atmosphere maintained during MBE deposition. The simulated STM image (Figure 4A) agrees excellently with the experimental counterpart (Figure 4B), both revealing a well-defined Kagome lattice. We note that the simulated STM contrast at the triangular bridge sites appears as bright features, whereas the corresponding regions in the experimental image appear as voids. This discrepancy may arises from the tunneling matrix-element effects[43]. The underlying Cr layer undergoes slight lattice distortion, forming a periodic hexagonal scaffold to accommodate the Sb surface reconstruction (Figure S17). The energy-dependence of the simulated STM images (Figure S18) also confirms the structural origin for the kagome-like morphology. Figure 4C presents the calculated Fermi surface comprises both electron-like and hole-like bands. The band structures of bilayer CrSb with spin-orbit coupling (SOC) are shown in Figure 4D, where the bands crossing the Fermi level are predominantly derived from Cr 3*d* orbitals (also Figures S19-S20).

Significantly, the electron depletion at the CrSb side and accumulation at the STO side indicate electron transfer from CrSb film to STO substrate (Figure 4E). The interfacial charge transfer predominantly occurs through hybridization between the Ti 3*d* orbitals of the STO substrate and the Cr atoms of the CrSb film, which will be discussed later. Specifically, sites A, B, and C correspond to the reconstructed Sb, Cr, and Sb atomic sites (Figure 2B), respectively, with site A being the farthest from the interface. As a result, site A experiences the weakest interfacial coupling and exhibits negligible spectral shift. In contrast, site B shows the most pronounced effect due to its direct interfacial hybridization, while site C exhibits a moderate change. This site-dependent behavior is fully consistent with the proposed charge-transfer picture.

Next, we discuss the emergence of ferromagnetic ordering in bilayer CrSb/STO films. The calculated magnetic moment of 1BL CrSb is 3.3 $\mu_B$ per Cr atom, consistent with the experimental value estimated from saturation magnetization 3.0 $\mu_B$ (Supplementary Note 3). We notice the calculated magnetic moment of bulk CrSb is 2.9 $\mu_B$ per Cr atom, indicating that dimensional reduction only slightly modulates the local magnetic moment of Cr atom. Accordingly, we infer that the emergence of ferromagnetism in bilayer CrSb, as opposed to the altermagnetic ground state in bulk CrSb, originates predominantly from the dimension-driven orbital reordering of Cr 3*d* states. As shown in Figure 5A, Cr atoms in bulk CrSb reside in a trigonal antiprismatic crystal field with $D_{3d}$ symmetry, whereas surface-exposed or top Cr atoms in 1BL CrSb experience a $D_{3h}$ crystal field. Compared to bulk CrSb, the Cr 3*d* orbitals in 1BL CrSb become more localized. Specifically, the $d_{xz}/d_{yz}$ and $dz^2$ orbitals in the spin-up channel shift toward the Fermi level and are preferentially occupied (Figures 5C & 5E). In bulk CrSb, coupling between half-occupied Cr $d_{xz}/d_{yz}$ orbitals in the spin-up channel favors antiferromagnetic exchange (Figures 5B & 5D). In contrast, in 1BL CrSb, the orbital reorganization gives rise to a competing FM exchange pathway involving occupied Cr $dz^2$ orbitals (spin-up state) and unoccupied Cr $dz^2$ orbitals (spin-down state), which becomes dominant over the AFM interaction. When all surface Cr atoms are capped by Sb atoms, the coordination environment of surface Cr atoms is restored to the $D_{3d}$ crystal field, and the magnetic ground state reverts to antiferromagnetism. As the film thickness increases, the electronic structure gradually recovers its bulk-like character, restoring the balance of exchange interactions in favor of antiferromagnetism.

The DFT calculations further reveal that the magnetic anisotropy energy (MAE) of bulk and BL CrSb is 0.01 and 0.32 meV per Cr atom, respectively, with the easy axis aligned along the out-of-plane direction for both (Supplementary Note 3). The enhanced MAE in BL CrSb is associated with the rearrangement of Cr 3*d* orbitals near the Fermi level due to reduced dimensionality (Figure S21). With the STO substrate considered, the total MAE of BL CrSb/STO further increases to 0.47 meV per Cr atom. The calculated value of $D(E_F)\times I$ is 4.41, which still satisfies the Stoner criterion (i.e., $D(E_F)\times I > 1$), confirming the itinerant ferromagnetic nature of 1BL CrSb/STO films. Within the framework of the mean-field Ising model, the Curie temperature ($T_c$) is predicted to be 347 K (Table S1), which is in excellent agreement with our experimental observation that the $T_c$ exceeds 300 K.

**DISCUSSION**

We discuss the interfacial effects in bilayer CrSb/STO heterostructures. First, the interfacial charge transfer, which is revealed by DFT calculations and experimentally verified via the rigid shift of dI/dV spectra in Figure 2, indicates the presence of strong interfacial interactions. Moreover, the evolution of the growth mode from layer-by-layer growth to island growth in bilayer and thicker CrSb films further confirms that the interfacial interaction is significantly stronger between bilayer CrSb films and the STO substrate. The observed thickness-dependent ferromagnetism in CrSb/STO resembles the interfacial doping effect reported for FeSe/STO[44], where charge transfer from the substrate is most effective for the first monolayer and rapidly diminishes with increasing thickness. In our case, the substrate-induced orbital reconstruction and enhanced MAE are similarly confined to the 1 BL regime, as supported by the calculated interface interaction energies of –0.11 eV/$Å^2$ (1 BL) versus –0.02 eV/$Å^2$ (2 BL). The rapid attenuation of the interfacial interaction with increasing thickness arises from the short-range nature of the Ti 3*d*–Cr orbital hybridization, which is screened by the intervening CrSb layers. This is reflected in the calculated interface interaction energies: –0.11 eV/Å$^2$ for 1 BL versus only –0.02 eV/Å$^2$ for 2 BL, indicating that the substrate effect is primarily confined to the first bilayer. The magnetization suppression observed in thicker CrSb films is attributed to a ferromagnetic-to-antiferromagnetic crossover, as supported by DFT calculations showing that both the interface interaction energy and the ferromagnetic exchange couplings decay rapidly with increasing thickness, allowing the bulk-like antiferromagnetic coupling to prevail (Supplementary Note 4).

Another evitable effect is the strain (Supplementary Note 1), which is expected to modify the crystal environment and strongly affect the spin-split electronic bands. We show that the band structure of 1BL CrSb can be effectively changed by strain (Figure S22). Previously, Song et al. reported a spontaneous anomalous Hall effect in ~30 nm CrSb films and attributed its origin to strain-induced crystal distortion by the substrate, which reduces the magnetic symmetry of the films[27]. Here, we provided experimental evidences of the ferromagnetic ordering emerges at the 2D limit in a bulk altermagnet CrSb. The existing uncovered Cr surface appears to be highly correlated with this magnetic transition. The observed perpendicular magnetic anisotropy may plays a crucial role in stabilizing magnetic ordering in 2D system, which supports the existence of 2D magnetism according to the Mermin-Wagner theorem[45]. Moreover, the observed Kagome lattice may intrigue quantum states arising from the interplay of spin-orbit coupling, electron correlations, and topology[46,47], and thus providing distinctive electronic properties in CrSb films. Nevertheless, the underlying mechanism for the magnetism in ultrathin CrSb films deserves further investigation.

In summary, we achieved the uniform bilayer CrSb films directly grown on STO(001) substrates and revealed its room-temperature ferromagnetism and high perpendicular magnetic anisotropy. Cr-terminated surfaces driving orbital rearrangements and interfacial coupling-enhanced MAE are the primary origins of robust ferromagnetism in ultrathin CrSb films. These results not only illuminate the vanishing altermagnetism and emergent ferromagnetism at the 2D limit of CrSb but also offer significant opportunities for exploring emergent quantum phenomena arising from the interplay between ferromagnetism and interfacial interactions, with promising implications for spintronics applications.

## METHODS

*Molecular Beam Epitaxy*. Our experiments were performed in an Omicron LT-STM system equipped with a home-made MBE chamber under a base pressure better than $2 \times 10^{-10}$ mbar. During the MBE growth, high purity Cr (99.995%) and Sb (99.9999%) elements were evaporated from standard K-cells onto Nb-doped (0.05 wt%) STO(001) substrates with size of 10 mm × 2 mm × 0.5 mm at temperatures between 300 and 400°C. The growth rate is 0.06 BL/min with an estimated Sb/Cr ratio 5:1. The epitaxial CrSb films on STO(001) substrates used in our experiments possess an effective area of 6 mm × 2 mm.

*Scanning Tunneling Microscopy/Spectroscopy*. The STM/S experiments were conducted at $T$ = 4.5 and 77 K with a polycrystalline Tungsten tip calibrated on Ag/Si(111) samples. STM topographic images were acquired in a constant current mode. Tunneling spectra were measured by disabling the feedback circuit, sweeping the sample bias $V$, and then extracting the differential conductance dI/dV with a standard lock-in technique under a small bias modulation of 20 meV at 987.5 Hz, unless otherwise specified.

*Ex-situ electrical and magnetization measurements*. Electrical transport and magnetic characterizations were measured by Physical Property Measurement System (PPMS) from Quantum Design integrated with a Vibrating Sample Magnetometer (VSM). As-grown CrSb films were capped with ~20 nm amorphous Si and Sb layers at a substrate temperature of ~100 K, and then transferred out of the vacuum chamber for *ex-situ* electrical and magnetization measurements, respectively. In-plane four-point probe electrical transport measurements were performed using silver paint and Al wire contacts. For magnetic measurements, the sample size was confined to 4 mm × 4 mm. We manually cut the samples to the required dimensions using a diamond scribe. During this process, the effective film area may vary slightly from sample to sample. To preclude potential magnetic contamination, all samples were manipulated exclusively with nylon tweezers. Magnetization measurements were performed at temperatures ranging from 2 to 300 K, over a magnetic field interval of –3.0 to 3.0 T and at a sweep rate of 50 Oe/s.

*Transmission Electron Microscopy measurements*. The TEM samples were prepared by a lift-out method in a Thermo Fisher Helios G4 UX focused ion beam (FIB) instrument at room temperature, and imaged in the JEOL JEM-1400 Flash TEM operating at 120 kV.

*Density-functional theory (DFT) calculations*. DFT calculations for CrSb were performed using the VASP by employing the Perdew-Burke-Ernzerhof (PBE) generalized gradient approximation (GGA) for exchange-correlation functional[48]. The projector-augmented-wave potential was used to describe the ion–electron interactions, and the energy cutoff of the plane waves was 500 eV.

## RESOURCE AVAILABILITY

### Lead contact

Requests for further information and resources should be directed to and will be fulfilled by the lead contact, Huimin Zhang (huiminzhang@dlut.edu.cn).

### Data availability

All raw data generated during the current study are available from the corresponding author upon request.

## ACKNOWLEDGMENTS

We thank Junwei Liu for insightful discussions. This work is supported by the National Key Research and Development Program of China (2023YFB3809600), National Natural Science Foundation of China (Grant No. 12304210, T2495211, 12274050 and 12534012), Young Top Talents in "Xingliao Talent Program" of Liaoning province (XLYC2403120), Natural Science Foundation of Liaoning Province (2026-MS-014), Open Research Fund Program of the State Key Laboratory of Low-Dimensional Quantum Physics, the Xiaomi Young Talents Program, and the Guangdong Provincial Quantum Science Strategic Initiative (GDZX2401002). We are grateful to the Instrumental Analysis Center, Dalian University of Technology for the technical supports.

## AUTHOR CONTRIBUTIONS

H.Z. conceived the research. Y.L., Z. W carried out the experiments. Q.L. and Y.Z. performed the DFT calculations. H.Z., X. J. and J. Z. analyzed the results and wrote the paper. All authors commented on the paper.

## DECLARATION OF INTERESTS

The authors declare no competing interests.

## FIGURE TITLES AND LEGENDS

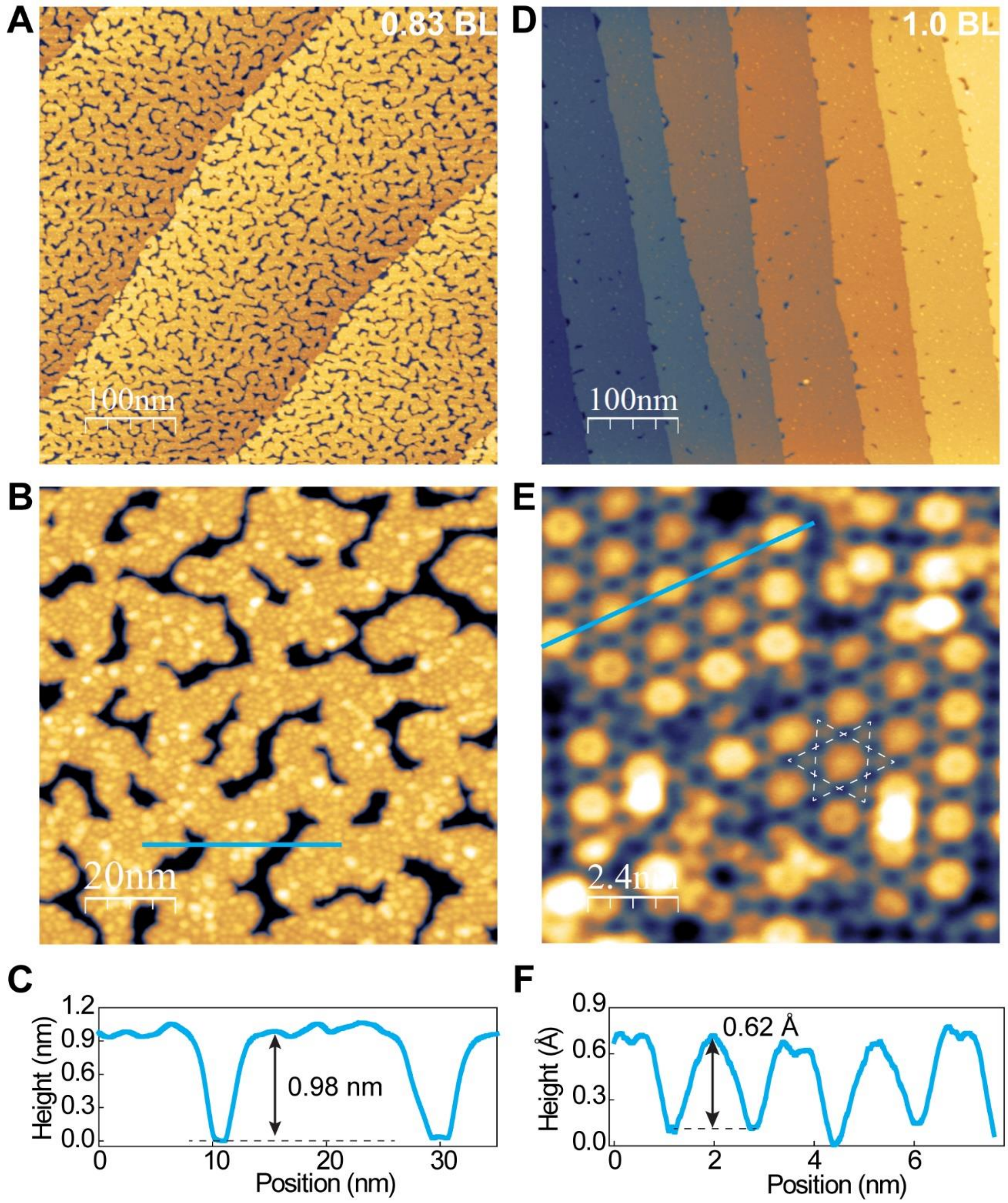

**Figure 1**. MBE-grown CrSb films on STO(001) substrate. (A) Topographic STM image of 0.83 BL CrSb/STO(001) films. Setpoint: $V$ = 1.0 V, $I$ = 1.0 nA. (B) Zoom-in image of (A). Setpoint: $V$ = 1.5 V, $I$ = 1.0 nA. (C) Line profile across the cyan line in (B). (D) Topographic STM image of 1.0 BL CrSb/STO(001) films. Setpoint: $V$ = 2.0 V, $I$ = 1.0 nA. (E) Atomic resolution image of the CrSb film. Setpoint: $V$ = 2.0 V, $I$ = 1.5 nA. The dashed white line denotes the Kagome lattice. (F) Line profile across the cyan line in (E).

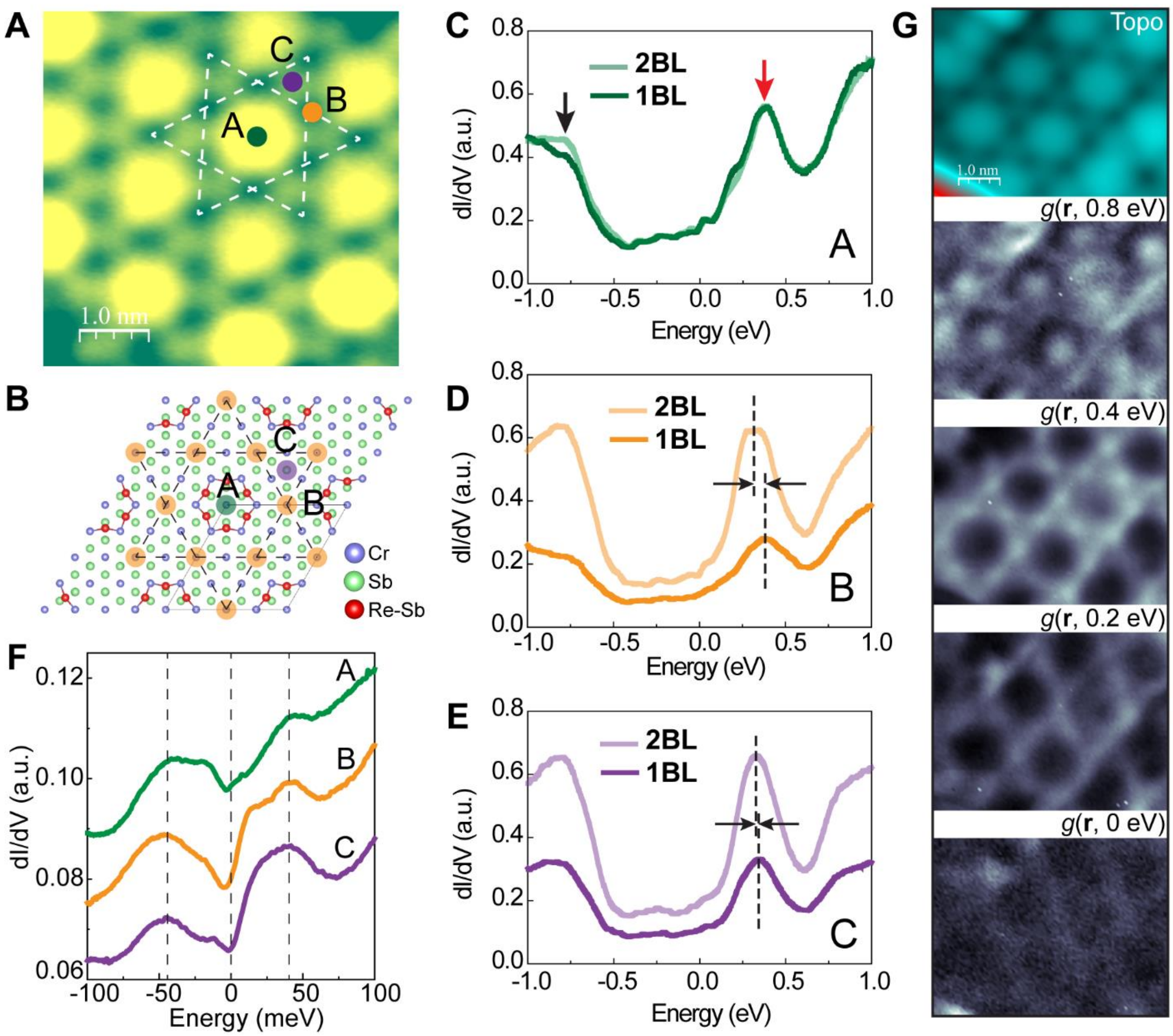

**Figure 2.** dI/dV spectra measurements of 1BL CrSb/STO films. (A) Topographic STM image revealing the Kagome lattice. Setpoint: $V = 0.2$ V, $I = 1.5$ nA. (B) Crystalline structure with three sites denoted as A, B and C in green, orange and purple, respectively. (C-E) Typical dI/dV spectra taken at the A, B and C sites marked in (A). (F) Zoom-in of the dI/dV spectra near the Fermi level for 1BL CrSb/STO. (G) dI/dV maps of the kagome lattice at various energies. Setpoint: $V = 2.0$ V, $I = 1.5$ nA, $V_{mod} = 40$ meV.

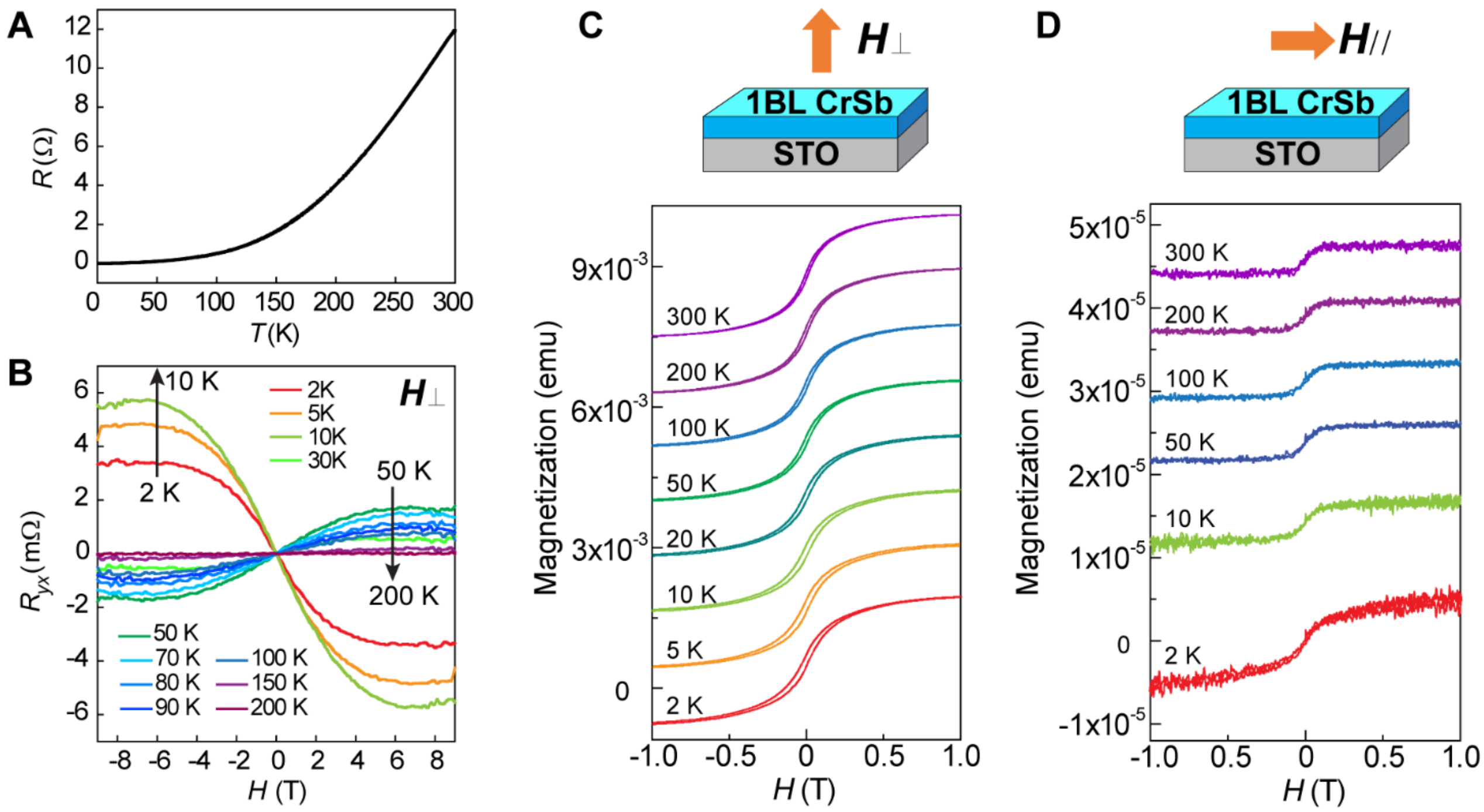

**Figure 3.** Electrical and magnetic properties of the 1BL CrSb/STO films. (A) Temperature-dependent resistance of the CrSb film grown on an insulating STO substrate and capped with an ~20 nm Si protective layer. (B) Hall resistance $R_{xy}$ as a function of perpendicular magnetic field ($H_{\perp}$) measured at various temperatures ranging from 2 to 200 K. (C-D) Magnetic field dependence of the magnetic moment (*M*-*H*) measured at 2, 5, 10, 20, 50, 100, 200, 300 K with magnetic field perpendicular (C) or parallel (D) to the film plane, respectively.

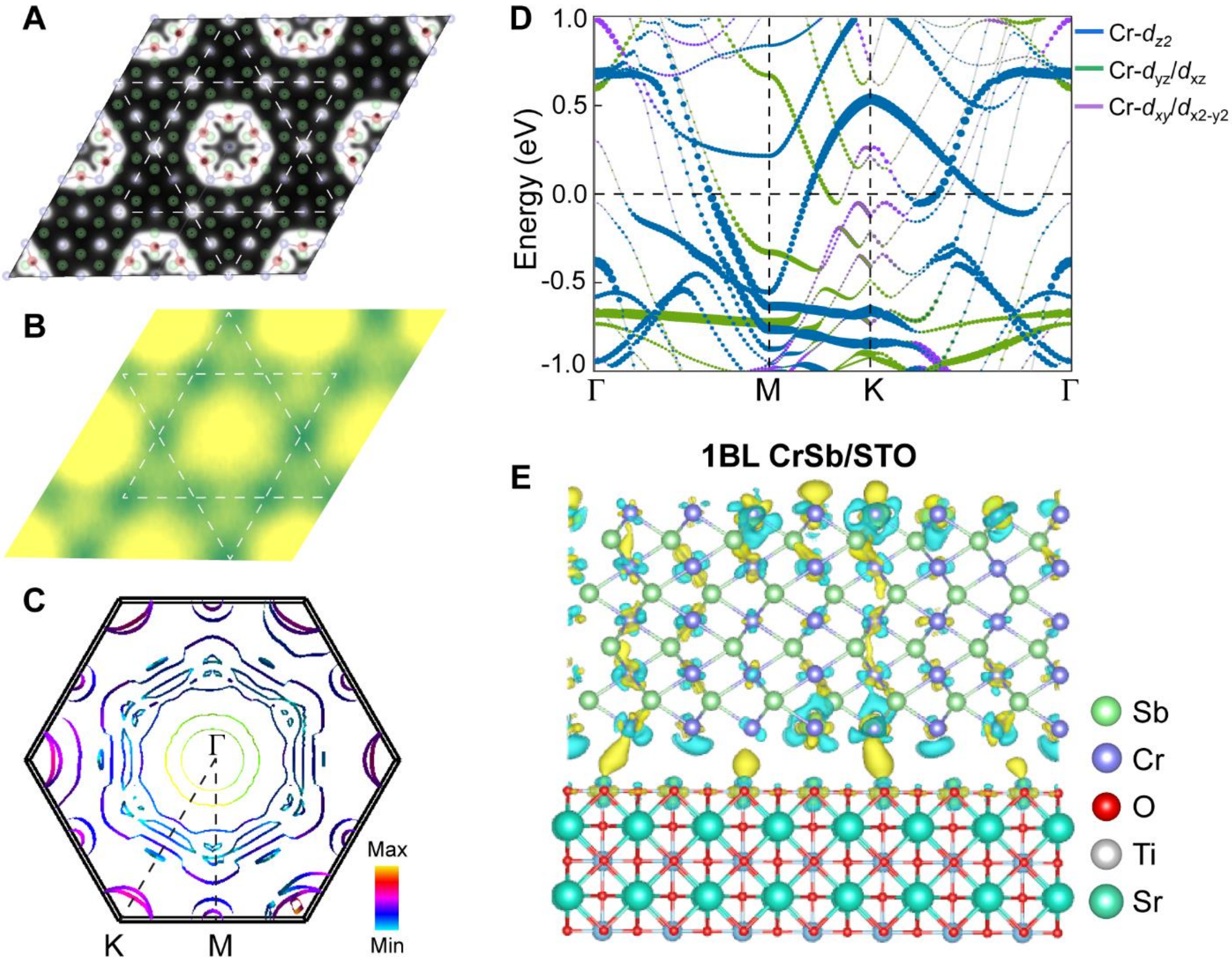

**Figure 4.** DFT calculations of 1BL CrSb/STO films. (A) DFT simulated STM image of the Cr terminated bilayer CrSb film with Sb reconstruction at $E$ = 0.1 eV (with crystal structure overlapped on top). (B) Experimental STM image showing the surface LDOS distribution. Setpoint: $V$ = 0.2 V, $I$ = 1.5 nA. (C) Calculated Fermi surface with SOC, where distinct colors represent different electron carrier densities. (D) Calculated band structure of 1BL CrSb with spin-orbit coupling (SOC). (E) Calculated charge density difference of 1BL CrSb/STO film (isovalue: 0.0005 e/Å$^3$). Blue and yellow represent charge depletion and accumulation, respectively. Right panel: the corresponding electrostatic potential profile along the vertical direction.

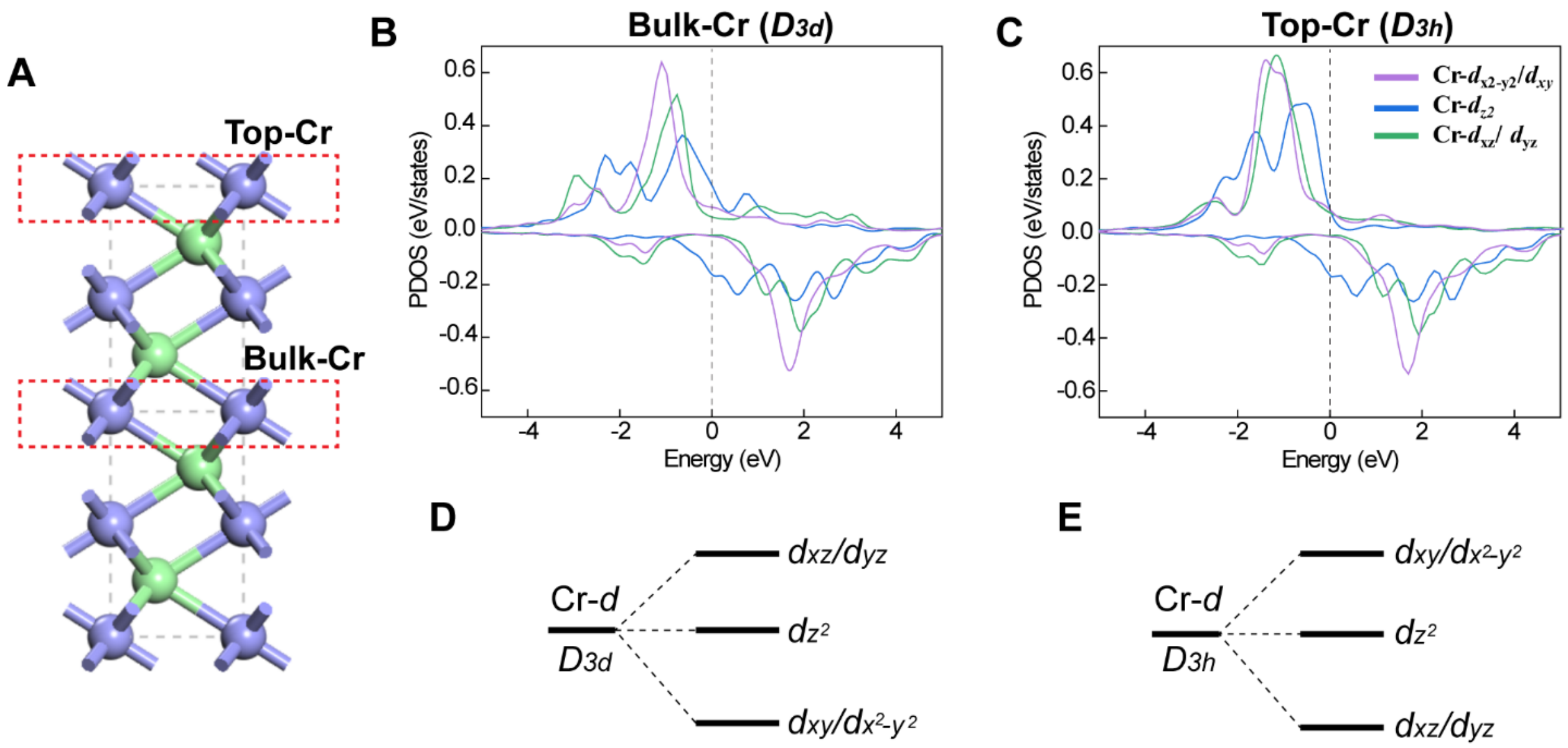

**Figure 5.** Magnetic properties of bilayer CrSb. (A) Site-dependent Cr atoms in CrSb films. (B-C) Calculated density of states contributed by Cr-3*d* orbitals of Cr atoms in the bulk (B) and top (C). (D-E) Schematic diagram of splitting of Cr-3*d* orbitals under $D_{3d}$ (D) and $D_{3h}$ (E) crystal field.

## SUPPLEMENTAL INFORMATION

Supplemental information can be found online at. Document S1. Supplemental Figures S1–S22